\pdfoutput=1

\documentclass[11pt,twoside,a4paper,cmspaper,final,collab]{cms-tdr}
\def\svnVersion{492978P}\def\cmsCernNoTag{CERN-EP-2018-312}\def\cmsCernDate{\today}\def\cmsMessage{Published in the Journal of High Energy Physics as \href{http://dx.doi.org/10.1007/JHEP03(2019)101}{\doi{10.1007/JHEP03(2019)101.}}}

\begin{document}\cmsNoteHeader{SUS-18-003}

\hyphenation{had-ron-i-za-tion}
\hyphenation{cal-or-i-me-ter}
\hyphenation{de-vices}
\RCS$HeadURL: svn+ssh://svn.cern.ch/reps/tdr2/papers/SUS-18-003/trunk/SUS-18-003.tex $
\RCS$Id: SUS-18-003.tex 492954 2019-03-28 13:30:25Z jrgonzal $

\newlength\cmsFigWidth
\ifthenelse{\boolean{cms@external}}{\setlength\cmsFigWidth{0.98\columnwidth}}{\setlength\cmsFigWidth{0.49\textwidth}}
\ifthenelse{\boolean{cms@external}}{\providecommand{\cmsLeft}{upper\xspace}}{\providecommand{\cmsLeft}{left\xspace}}
\ifthenelse{\boolean{cms@external}}{\providecommand{\cmsRight}{lower\xspace}}{\providecommand{\cmsRight}{right\xspace}}
\newlength\cmsTabSkip\setlength{\cmsTabSkip}{1ex}
\newcommand{\amcatnlo}   {\textsc{mg5}\_a\MCATNLO}
\newcommand{\mtll}{\ensuremath{M_{\textrm{T2}}}\xspace}

\newcommand{\Ttt}{\ensuremath{\text{T2}\PQt\PQt}\xspace}
\newcommand{\mtop}{\ensuremath{m_\PQt}\xspace}
\newcommand{\pp}{\ensuremath{\Pp\Pp}}
\newcommand{\tW}{\ensuremath{\cPqt\PW}}
\newcommand{\ttW}{\ensuremath{\cPqt\cPaqt\PW}}
\newcommand{\ttZ}{\ensuremath{\cPqt\cPaqt\PZ}}
\newcommand{\pchi}{\PSGczDo}
\newcommand{\pstop}{\PSQtDo}
\newcommand{\mchi}{\ensuremath{{m}_{\pchi}}}
\newcommand{\mstop}{\ensuremath{{m}_{\pstop}}}

\cmsNoteHeader{SUS-18-003}
\title{Search for the pair production of light top squarks in the $\Pe^{\pm}\Pgm^{\mp}$ final state in proton-proton collisions at $\sqrt{s} = 13$\TeV }

\date{\today}

\abstract{
A search for the production of a pair of top squarks at the LHC is presented. This search targets a region of parameter space where the kinematics of top squark pair production and top quark pair production are very similar, because of the mass difference between the top squark and the neutralino being close to the top quark mass. The search is performed with 35.9\fbinv of proton-proton collisions at a centre-of-mass energy of $\sqrt{s} = 13\TeV$, collected by the CMS detector in 2016, using events containing one electron-muon pair with opposite charge. The search is based on a precise estimate of the top quark pair background, and the use of the \mtll variable, which combines the transverse mass of each lepton and the missing transverse momentum. No excess of events is found over the standard model predictions. Exclusion limits are placed at 95\% confidence level on the production of top squarks up to masses of 208\GeV for models with a mass difference between the top squark and the lightest neutralino close to that of the top quark.
}

\hypersetup{
pdfauthor={CMS Collaboration},
pdftitle={Search for the pair production of light top squarks in the emu final state in proton-proton collisions at sqrt(s) = 13 TeV},
pdfsubject={CMS},
pdfkeywords={CMS, supersymmetry, squarks}}

\maketitle

\section{Introduction}

The standard model (SM) of particle physics accurately describes the vast majority of the observed particle physics phenomena.
However, there are several open problems that cannot be explained by the SM, such as the hierarchy problem, the need for fine tuning to explain the large difference between the electroweak and the Planck scale~\cite{BARBIERI198863, WITTEN1981513}, and the lack of a candidate particle that explains the nature of dark matter in cosmological and astrophysical observations~\cite{Bertone:2004pz, DarkMatter2}.
Supersymmetry (SUSY)~\cite{Ramond:1971gb, Golfand:1971iw, Neveu:1971rx, Volkov:1972jx, Wess:1973kz, Wess:1974tw, Fayet:1974pd, Nilles:1983ge, Martin:1997ns} is a well-motivated extension
of the SM that provides a technically natural~\cite{Hierarchy1, DIMOPOULOS1981150} solution to both of these problems, through the introduction of an additional symmetry between bosons and fermions. In SUSY models, large quantum loop corrections to the masses of the Higgs bosons, mainly produced by the top quark, are mostly cancelled by the one produced by its SUSY partner, the top squark (\pstop), if their masses are close in value. Similar cancellations occur for other particles, resulting in a natural solution to the hierarchy problem. Furthermore, SUSY introduces a new quantum number, R-parity~\cite{Farrar:1978xj}, that distinguishes between SUSY and SM particles. If R-parity is conserved~\cite{Farrar:1978xj}, top squarks are produced in pairs and the lightest SUSY particle is stable, which if neutral (\pchi) provides a good candidate for dark matter.
The lighter SUSY particles may have masses close to those of the SM particles, and therefore could be produced in proton-proton (\pp) collisions within the energy reach of the CERN LHC. In certain scenarios the lightest top squarks are expected to have a mass (\mstop) close to the top quark mass (\mtop), leading to a natural solution to the hierarchy problem~\cite{Hierarchy1,DIMOPOULOS1981150,Czakon:2014fka}.

This paper presents a search for the production of a pair of scalar top partners and neutralinos that are degenerate or nearly degenerate in mass with the top quark ($\mstop - \mchi\simeq \mtop$), using events
produced in \Pp\Pp~collisions at a centre-of-mass energy of 13\TeV recorded with the CMS detector at the LHC.
A data sample collected during 2016 and corresponding to an integrated luminosity of 35.9\fbinv is used.

\begin{figure}[htb!]
   \centering
     \includegraphics[width=1\cmsFigWidth]{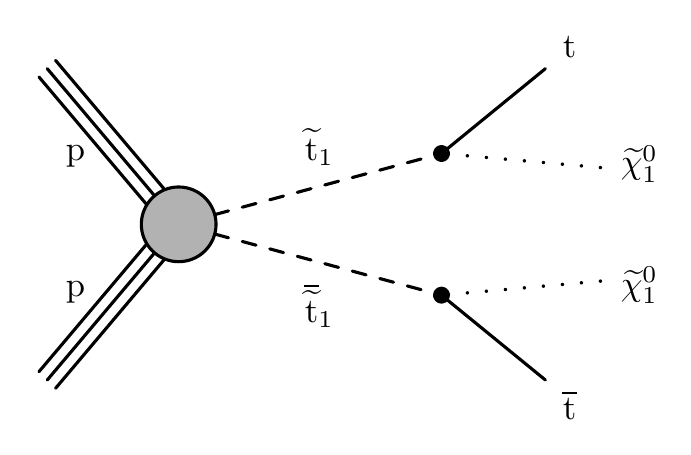}
     \caption{Diagram of the top squark pair production with further decay into a top (antitop) quark and the lightest neutralino.}
     \label{fig:T2tt}

 \end{figure}

Top squarks in this search are assumed to decay as $\pstop\to\cPqt\pchi$, as shown in Fig.~\ref{fig:T2tt}.
In particular, this analysis uses events in which the resulting top (anti)quark decays into a bottom (anti)quark and a \PW~boson that in turn decays into a lepton and a neutrino, and selects final states characterized by the presence of an opposite-sign electron-muon pair.

Given that the target SUSY signal and the SM top quark pair (\ttbar) production processes are characterized by equivalent final states with very similar kinematics, most of the top squark searches by the ATLAS~\cite{Aaboud:2017nfd,Aad:2015pfx,Aad:2014kra,Aad:2014qaa,Aaboud:2016lwz}
and CMS~\cite{Sirunyan:2017xse,Sirunyan:2017leh,Chatrchyan:2013xna,Khachatryan:2016pup,Khachatryan:2016pxa,Sirunyan:2016jpr,Sirunyan:2017wif,Sirunyan:2017pjw} Collaborations do not have enough sensitivity for observing the production of top squarks in these scenarios.
Limits on the production cross section of signals described by these models have previously been set through \ttbar production cross section measurements at 8\TeV by the CMS~\cite{Khachatryan:2016mqs} and ATLAS~\cite{Aad:2014kva,Aad:2014mfk} Collaborations, excluding the presence of a top squark with a mass of up to 191\GeV for a neutralino mass of 1\GeV.

The analysis is performed as a search for an excess above a large \ttbar background, which must be estimated precisely to attain sensitivity to the signal. Further separation is achieved by exploiting the distribution of signal and background events in a discriminating variable (\mtll).

\section{The CMS detector}
The central feature of the CMS apparatus is a superconducting solenoid of 6\unit{m} internal diameter, providing a magnetic field of 3.8\unit{T}. Within the solenoid volume are a silicon pixel and strip tracker covering the full range of the azimuthal angle $0 < \varphi < 2\pi$ and a pseudora\-pidity of $\abs{\eta} < 2.5$, a lead tungstate crystal electromagnetic calorimeter (ECAL), and a brass and scintillator hadron calorimeter (HCAL), each composed of a barrel and two endcap sections. Forward calorimeters extend the pseudorapidity coverage provided by the barrel and endcap detectors. Muons are detected in gas-ionization chambers embedded in the steel flux-return yoke outside the solenoid.

Events of interest are selected using a two-tiered trigger system~\cite{Khachatryan:2016bia}. The first level, composed of custom hardware processors, uses information from the calorimeters and muon detectors to select events at a rate of around 100\unit{kHz} within a time interval of less than 4\mus. The second level, known as the high-level trigger, consists of a farm of processors running a version of the full event reconstruction software optimized for fast processing, and reduces the event rate to around 1\unit{kHz} before data storage.

A more detailed description of the CMS detector, together with a definition of the coordinate system used and the relevant kinematic variables, can be found in Ref.~\cite{bib:cms}.

\section{Monte Carlo simulation}\label{sec:MC}
A correct estimate of the \ttbar background is crucial for this analysis and the uncertainties on the modelling of this process plays an important role, especially the theoretical uncertainties on the \ttbar cross section.

The \POWHEG v2~\cite{bib:powheg2,Frixione:2007vw,Nason:2004rx} generator is used to simulate \ttbar events at the next-to-leading order (NLO) in quantum chromodynamics (QCD), as well as to calculate the dependency of the \ttbar acceptance on \mtop, and on the factorization ($\mu_\mathrm{F}$) and renormalization ($\mu_\mathrm{R}$) scales.
A parameter, denoted as damping factor $h_{\mathrm{damp}}$, is used to limit the resummation of higher-order effects by the Sudakov form factor to below a given transverse momentum (\pt) scale~\cite{CMS-PAS-TOP-16-021}. The central value and uncertainties of $h_{\mathrm{damp}}$ will be discussed later.

Single top quark and antiquark production in association with a \PW~boson (\tW) is simulated at NLO using the \POWHEG v1~\cite{bib:powheg3} generator.
The Drell--Yan process (DY), and the production of \PW\ or \cPZ\ bosons in association with \ttbar events (referred to as {\ttbar}V), are generated at NLO using the \amcatnlo~v2.2.2~\cite{Alwall:2014hca} generator. The production of the DY process is simulated with up to two additional partons and the FxFx scheme is used for the matching of the matrix elements and parton showers~\cite{Frederix:2012ps}. The contributions from \PW\PW, \PW\cPZ, and \cPZ\cPZ\ (collectively referred to as VV) processes are simulated at leading order (LO) using \PYTHIA~v8.205~\cite{Sjostrand:2014zea}.

The \Ttt model from the simplified model spectra~\cite{Alwall:2008ag, Alves:2011wf} is used to model the SUSY signal, in which top quarks are unpolarized and a branching fraction of 100\% is assumed for the top squark decaying into a top quark and a neutralino. The generation of signal samples is performed using the \amcatnlo~generator at LO.

The NNPDF 3.0~\cite{Ball:2014uwa} parton distribution function (PDF) set is used for all the samples.
Parton showering and hadronization are handled by \PYTHIA using the underlying event tune CUETP8M2T4~\cite{CMS-PAS-TOP-16-021} for SM \ttbar~events and the CUETP8M1~\cite{Khachatryan:2015pea} tune for all other background and signal events.

The response of the CMS detector is simulated for all the generated events with the \GEANTfour package~\cite{Agostinelli:2002hh}.
The effect of additional interactions in the same events (referred to as pileup) is accounted for by simulating additional interactions for each hard scattering event. Simulated events are then reweighted so that the simulated pileup vertex distribution matches the observed distribution, which has an average of 23 collisions per bunch crossing.

Simulated events are normalized according to the integrated luminosity and the theoretical cross section of each process.
The latter are computed at next-to-next-to-leading order (NNLO) (DY~\cite{PhysRevD.86.094034}),
approximate NNLO order (\tW~\cite{Kidonakis:2015nna}), and NLO (VV~\cite{bib:mcfm:diboson}, {\ttbar}V~\cite{Garzelli:2012bn}).

For the normalization of the simulated \ttbar sample, the full NNLO plus next-to-next-to-leading-logarithmic accuracy calculation~\cite{Czakon:2013goa} is used, performed with the \textsc{Top++} 2.0 program~\cite{Czakon:2011xx}. The PDF uncertainties are added in quadrature
to the uncertainty associated with the strong coupling constant (\alpS) to obtain a \ttbar production cross section of $832~^{+20}_{-29}\,\text{(scale)}\pm 35\,$(PDF+\alpS)\unit{pb} assuming $\mtop = 172.5\GeV$.

The signal events are normalized to the theoretical NLO cross section~\cite{Borschensky:2014cia,Beenakker:2011fu, Beenakker:2009ha, Kulesza:2009kq, Kulesza:2008jb, Beenakker:1996ch} obtained from the simplified model spectrum for the \Ttt model.

\section{Objects and event selection} \label{sec:selection}
In the SM, top quarks decay almost exclusively into a bottom quark and a \PW~boson. In this analysis, events containing an $\Pe^{\pm}\Pgm^{\mp}$ pair and jets are selected. Signal events may have a larger amount of missing transverse momentum (\ptmiss) with respect to \ttbar events because of the presence of the neutralinos.

Events are required to pass a dilepton trigger based on the presence of one electron (muon) with $\pt > 23\,(23)\GeV$ and one muon (electron) with $\pt > 8\,(12)\GeV$. To increase the trigger efficiency, events passing a single-lepton trigger that requires the presence of one electron (muon) with $\pt > 35\,(24)\GeV$ are also selected. The efficiency of the combination of dilepton and single-lepton triggers for events with an electron-muon pair with $\pt > 25$ and 20\GeV is measured in data and found to be approximately 98\%. The simulated trigger efficiency is corrected to match that observed in data by using a multiplicative scale factor calculated as a function of the pseudorapidity of the leptons.

The particle-flow (PF) algorithm~\cite{bib:PF} aims to reconstruct and identify each individual particle in an event, with an optimized combination of information from the various elements of the CMS detector. The reconstructed vertex with the largest value of summed physics object $\pt^2$ is taken to be the primary $\Pp\Pp$ interaction vertex, where the physics objects are the objects returned by a jet finding algorithm~\cite{Cacciari:2008gp, Cacciari:2011ma} applied to all charged tracks associated with the vertex, plus the corresponding associated \ptmiss. The energy of photons is obtained from the ECAL measurement. The energy of electrons is determined from a combination of the electron momentum at the primary interaction vertex as determined by the tracker, the energy of the corresponding ECAL cluster, and the energy sum of all bremsstrahlung photons spatially compatible with originating from the electron track. The momentum of muons is obtained from the curvature of the corresponding track. The energy of charged hadrons is determined from a combination of their momentum measured in the tracker and the matching ECAL and HCAL energy deposits, corrected for zero-suppression effects and for the response function of the calorimeters to hadronic showers. Finally, the energy of neutral hadrons is obtained from the corresponding corrected ECAL and HCAL energies.

Selected leptons (electrons and muons) are required to have $\pt \geq 20\GeV$, $\abs{\eta} \leq 2.4$, and to satisfy a lepton isolation criterion. The lepton isolation variable is defined as the scalar \pt sum of all the PF candidates inside a cone of $\Delta R = \sqrt{\smash[b]{(\Delta\eta)^2+(\Delta\phi)^2}} = 0.3\,(0.4)$ centered on the electron (muon) candidate, excluding the contribution from the lepton candidate itself. To account for particles produced in pileup interactions, the contribution from charged hadrons that are not associated to the primary vertex is removed and a correction is applied for the expected contribution of neutral hadrons, following the procedure in \cite{Khachatryan:2015hwa}. This isolation variable is required to be smaller than 6 (15)\% of the electron (muon) candidate \pt. Selected leptons are required to originate from the primary vertex.

Jets are reconstructed from PF candidates using the anti-\kt clustering algorithm~\cite{Cacciari:2008gp, Cacciari:2011ma} with a distance parameter of 0.4. The jet momentum is defined as the vector sum of the momenta of all PF candidates associated with the jet, and is found to be within 5--10\% of the true momentum over the entire \pt spectrum. The charged PF candidates that are determined to originate from pileup vertices are discarded in the jet reconstruction, and an offset correction is applied to account for remaining contributions of the pileup interactions \cite{Khachatryan:2016kdb}. Selected jets are required to have $\pt \geq 30\GeV$ and $\abs{\eta} \leq 2.4$ and must come from the main primary vertex. In order to avoid double counting, jets that overlap with the selected leptons in a cone of $\Delta R = 0.4$ are not considered.

Jets originating from \cPqb~quarks are identified (tagged) as \cPqb~jets using the combined secondary vertex algorithm v2~\cite{Sirunyan:2017ezt}. This algorithm combines the information of the reconstructed secondary vertex with other kinematic variables of the jet by using a multivariate classifier to maximize the probability of tagging \cPqb~jets. An operating point that yields identification efficiencies of about 70\% is used. The corresponding misidentification probabilities are about 1\% for light-flavour jets (originating from \cPqu, \cPqd, \cPqs~quarks or gluons) and 15\% for \cPqc~jets.

Lepton reconstruction, identification, and isolation efficiencies, as well as efficiencies for \cPqb~tagging and \cPqb~tag misidentification of light quarks or gluons are corrected in the Monte Carlo (MC) simulation to match the observed values. These corrections are parameterized as functions of the \pt and $\eta$ of the object and are of the order of 1\% for leptons and a few percent for jets~\cite{Sirunyan:2017ezt}.

The correction of MC efficiencies to match that observed does not introduce any bias in our search for an excess above SM background prediction as the lepton reconstruction, identification, and isolation efficiencies and the trigger efficiency are measured using the
tag-and-probe method \cite{Khachatryan:2015hwa,Sirunyan:2018fpa}, and \cPqb~tagging and \cPqb~tag misidentification rates are measured using an independent sample of QCD multijet events. In addition, these corrections are applied by bins of $\eta$ and \pt, the latter except for the trigger efficiency.

The vectorial missing transverse momentum (\ptvecmiss) is defined as the transverse component of the negative vector sum of the momenta of all reconstructed PF candidates in an event; its magnitude is denoted as \ptmiss. All the corrections applied to the jet momenta are propagated to the calculation of \ptmiss~\cite{bib:CMS-PAS-JME-16-004}.

Events containing one electron-muon pair with opposite charge and invariant mass greater than 20\GeV, to avoid selecting low mass resonances, are selected. The transverse momentum of the highest-\pt (leading) lepton must be at least 25\GeV. In case more than two leptons are present in the event, the dilepton pair is formed using the two highest \pt leptons, and the event is selected if that pair satisfies the aforementioned requirements. Selected events are also required to contain at least two jets, at least one of which must be a \cPqb-tagged jet.

\section{Search strategy}
After the event selection, the vast majority of events ($\approx$98\%) come from top quark production processes (\ttbar, \tW). For a top squark mass similar to that of the top quark, the production cross section of the signal process is expected to amount to up to 125\unit{pb}, corresponding to about 15\% of the SM \ttbar~production cross section. However, the kinematics of the final-state particles are very similar in both processes, so a control region for the \ttbar background with small signal contamination is impossible to define. The sensitivity of the analysis comes from a precise estimate of the \ttbar~background, using MC simulation and exploiting the 6\%~\cite{Czakon:2011xx} theoretical uncertainties on the predicted cross section and the even smaller \cite{Khachatryan:2016kzg, Khachatryan:2016mqs} experimental uncertainties on the measurement. Additional sensitivity comes from the small kinematic differences between the target signal and the \ttbar~background, which become more important with increasing top squark mass and increasing mass difference between the top squark and neutralino.

For a top squark mass of 245\GeV, the cross section decreases to $\approx$24\unit{pb}, but the presence of massive neutralinos ($\mchi > 50\GeV$) in the event can result in additional \ptmiss. To account for this, following previous top squark searches~\cite{Khachatryan:2016pup}, the sensitivity of the analysis is further increased by using the shape of the \mtll variable, defined as
\begin{equation}
\mtll = \min_{\vec{p}_{\textrm{T,1}}^{\textrm{miss}} + \vec{p}_{\textrm{T,2}}^{\textrm{miss}} = \ptvecmiss} \left( \max \left[ \mT(\vec{p}_\textrm{T}^{{\ell}1},\vec{p}_{\textrm{T,1}}^{\textrm{miss}}) , \mT(\vec{p}_\textrm{T}^{{\ell}2},\vec{p}_{\textrm{T,2}}^{\textrm{miss}}) \right] \right),
\end{equation}
where \mT is the transverse mass and $\vec{p}_\textrm{T1}^\textrm{miss}$, $\vec{p}_\textrm{T2}^\textrm{miss}$ correspond to the estimated transverse momenta of two neutrinos that are presumed to determine the total \ptvecmiss of the event. The transverse mass is calculated for each lepton-neutrino pair, for different assumptions of the neutrino \pt. The computation of \mtll is done using the algorithm discussed in Ref.~\cite{Cheng:2008hk}. The \mtll distribution has a kinematic endpoint at the mass of the \PW~boson in the case of \ttbar events~\cite{Lester:1999tx}, while this is not true if extra invisible particles are present in the event. For models where $\mstop \approx \mtop$, the discriminating power of \mtll is limited but the signal cross section is high enough to have sensitivity to the presence of a signal over the background expectation. Since events with $\mtll = 0\GeV$ do not provide any discrimination between signal and \ttbar background, only events with $\mtll > 0\GeV$ are used for hypothesis testing.

\begin{figure}[htb!]
   \centering
     \includegraphics[width=1\cmsFigWidth]{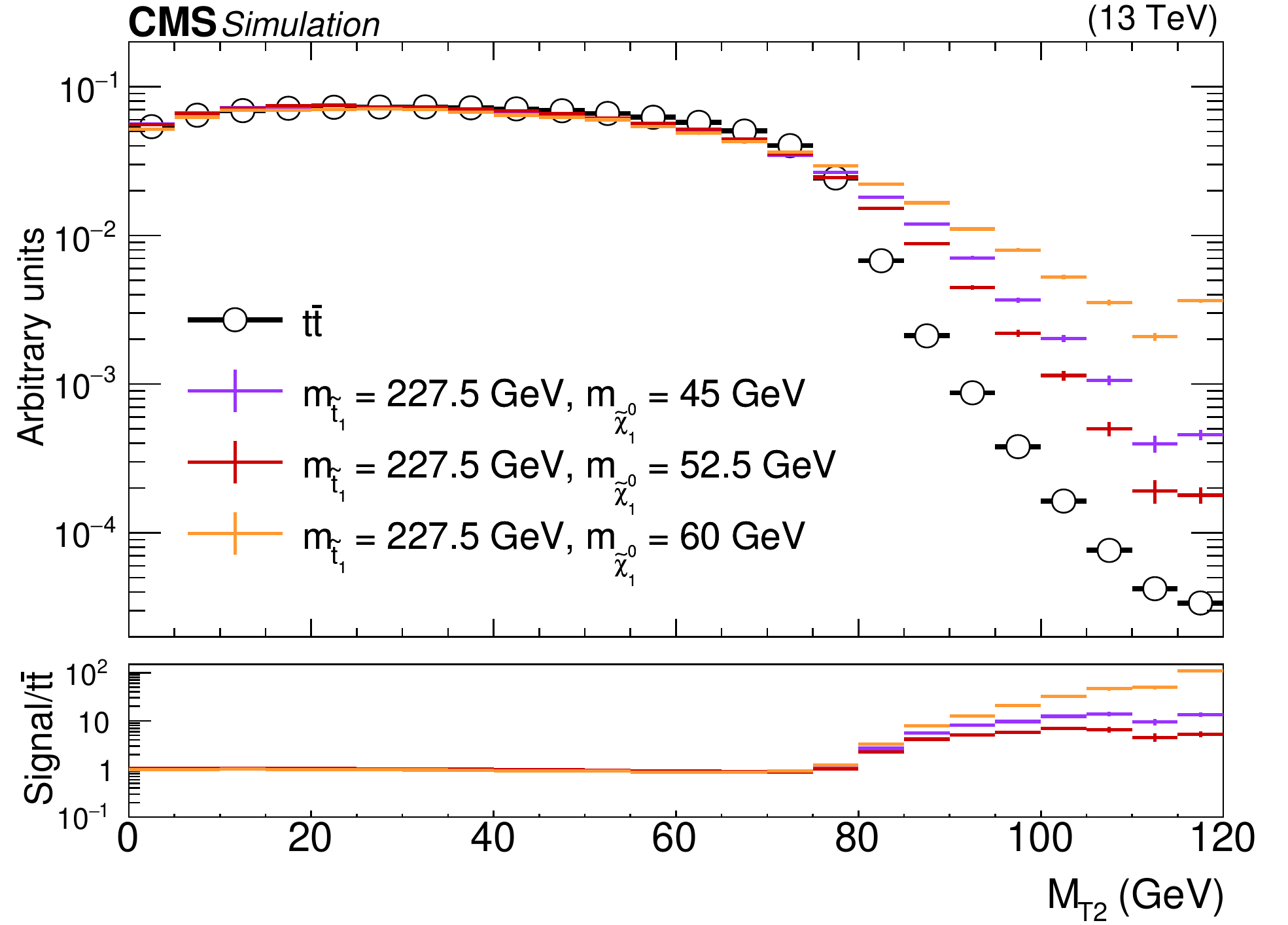}
     \includegraphics[width=1\cmsFigWidth]{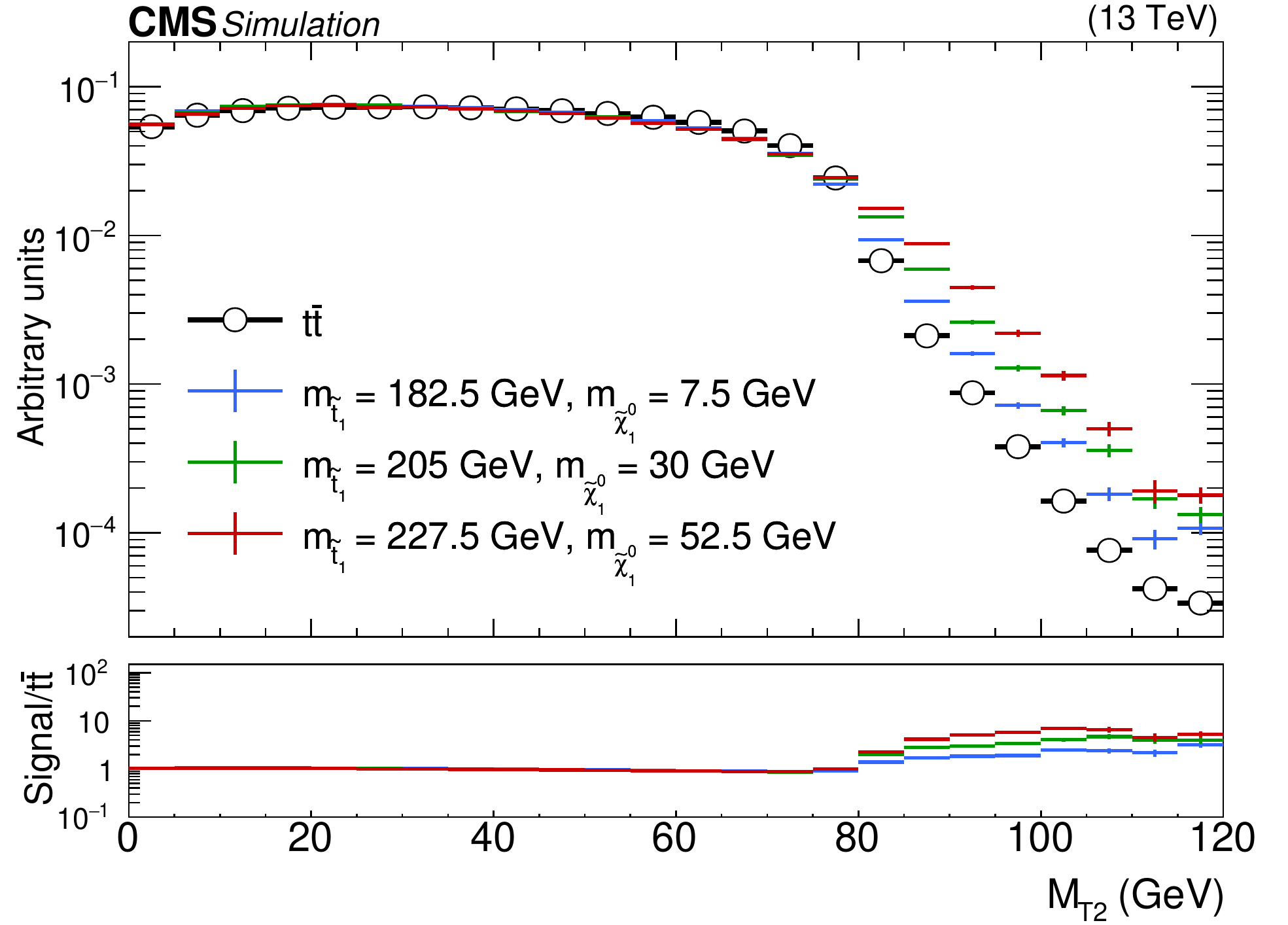}
     \caption{Normalized \mtll~distributions for various mass hypotheses for the top squark and for the neutralino. Variables at the generator level are used for \ttbar and signal events with two generated leptons with \pt of at least 20\GeV and $\abs{\eta} \leq 2.4$. The last bin includes the overflow.}
     \label{fig:MT2gen}

 \end{figure}

Figure~\ref{fig:MT2gen} shows the \mtll distributions for signal and background for different mass hypotheses for the stop squark and neutralino. The \mtll distributions of the simulated signal models are characterized by a slightly different shape for \mtll values smaller than 80\GeV and a large difference for $\mtll > 80\GeV$, because of the presence of the endpoint in the \mtll distribution for \ttbar events. This difference increases significantly when $\Delta m = \mstop - \mchi$ is different from the top quark mass (Fig.~\ref{fig:MT2gen} left). Furthermore, the differences in \mtll are large for signal points characterized by large neutralino masses, which have additional \ptmiss to the event (keeping $\Delta m \approx \mtop$, Fig. \ref{fig:MT2gen} right).

\section{Background estimation}
The \ttbar process accounts for approximately 94\% of the total background yields in the selected region, and is modelled from MC simulation using the sample described in Section \ref{sec:MC}. For this modelling, a top quark mass of 172.5\GeV is assumed. The accurate knowledge of the \ttbar production process has been previously demonstrated in several cross section measurements by the CMS Collaboration~\cite{Khachatryan:2016mqs}. Moreover, its differential cross section as a function of different variables has been measured ~\cite{Sirunyan:2018ptc} and MC parameters have been tuned using an independent data sample~\cite{CMS-PAS-TOP-16-021}. The MC tuning does not produce a significant modification of the \mtll shape. The main parameters affecting the \ttbar modelling and their associated uncertainties are discussed in Section \ref{sec:unc}. The \tW~background gives the second-largest contribution, approximately 4\%, and is also modelled using MC simulation.

The number of events with nonprompt leptons, including the contribution  of events with jets misidentified as leptons or with leptons coming from the decay of a bottom quark mistakenly identified as coming from the hard process, is estimated from an observed control region in which the electron and muon are required to have the same sign of the electric charge (referred to as \textit{same-sign}), while all other requirements for the event selection are the same as for the signal region. This background is estimated using the observed events in the control region after subtraction of the contribution from the backgrounds that produce prompt leptons. This contribution is estimated from MC simulation and comes mainly from \ttW~and \ttZ~events or dileptonic \ttbar with a mismeasurement of the electron charge. The events in this control region are weighted by the expected ratio of opposite-sign to same-sign events with nonprompt leptons after the full event selection, which is estimated in MC simulation to be $1.2\pm 0.1\syst$.

Other background contributions are estimated using MC simulation and come from DY, VV (\PW\PW, \PW\cPZ, and \cPZ\cPZ), \ttW, and \ttZ~events, for a total contribution of about 1\%.

\begin{figure}[htb!]
   \centering
     \includegraphics[width=1\cmsFigWidth]{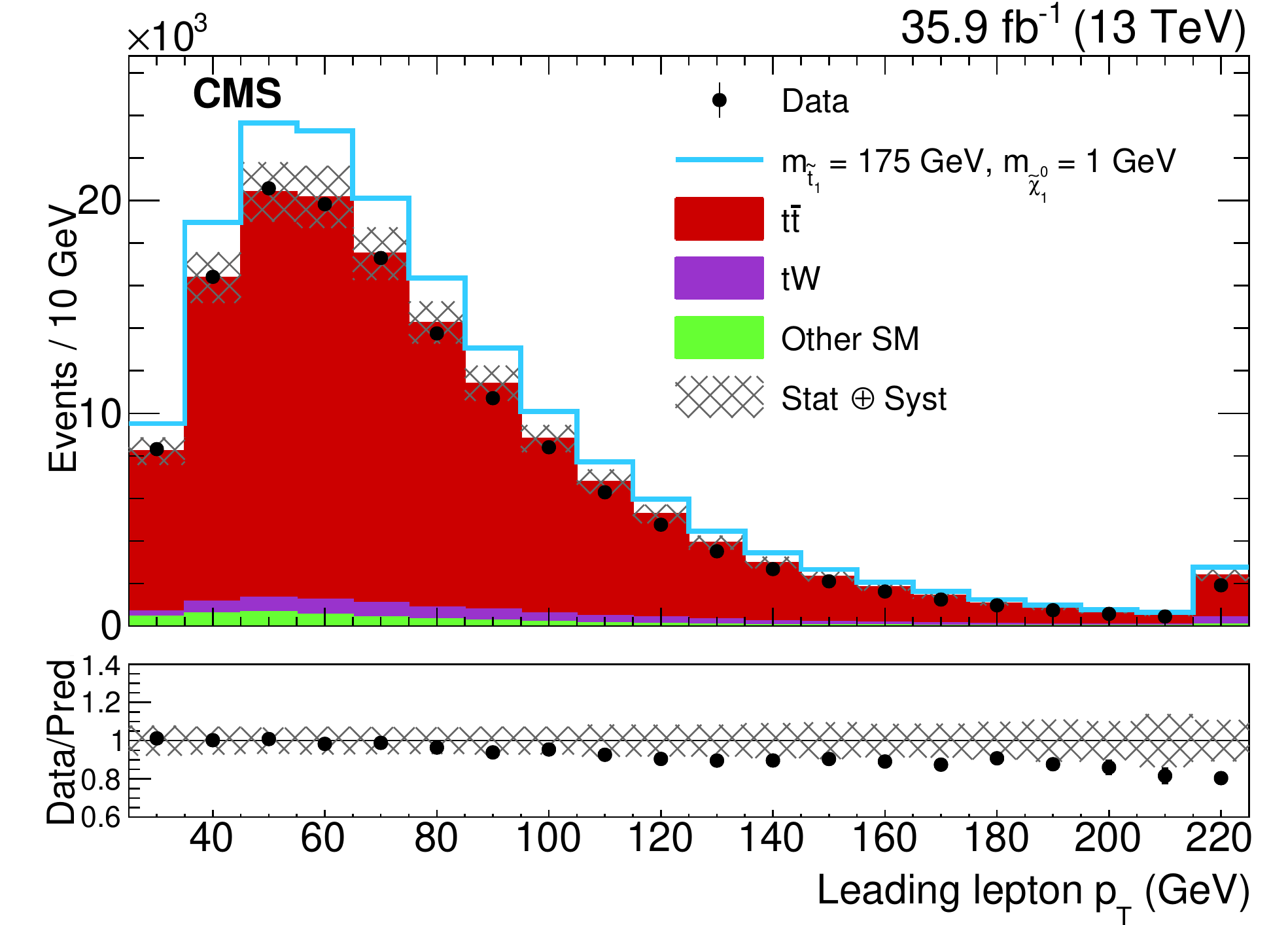}
     \includegraphics[width=1\cmsFigWidth]{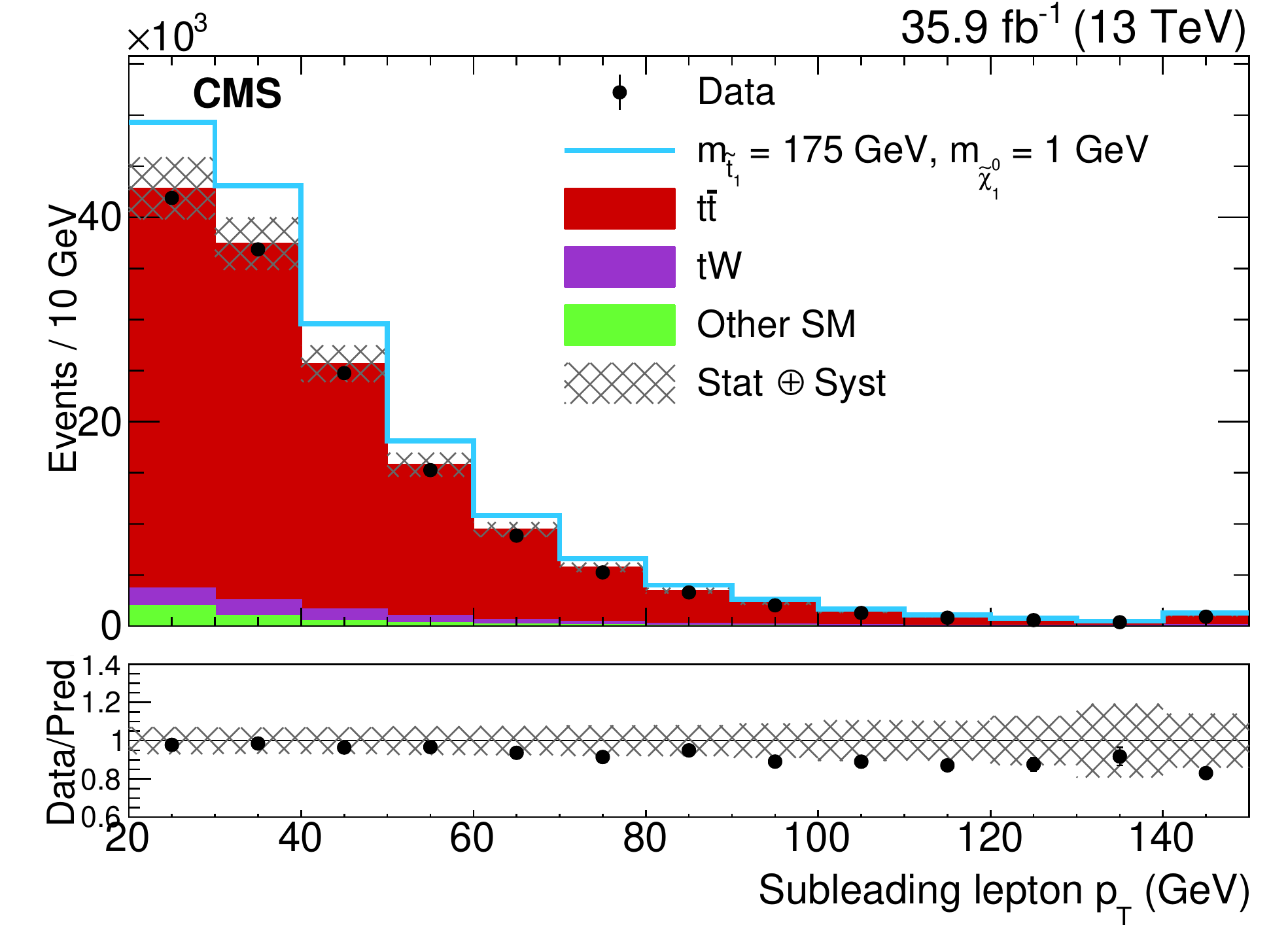}
     \includegraphics[width=1\cmsFigWidth]{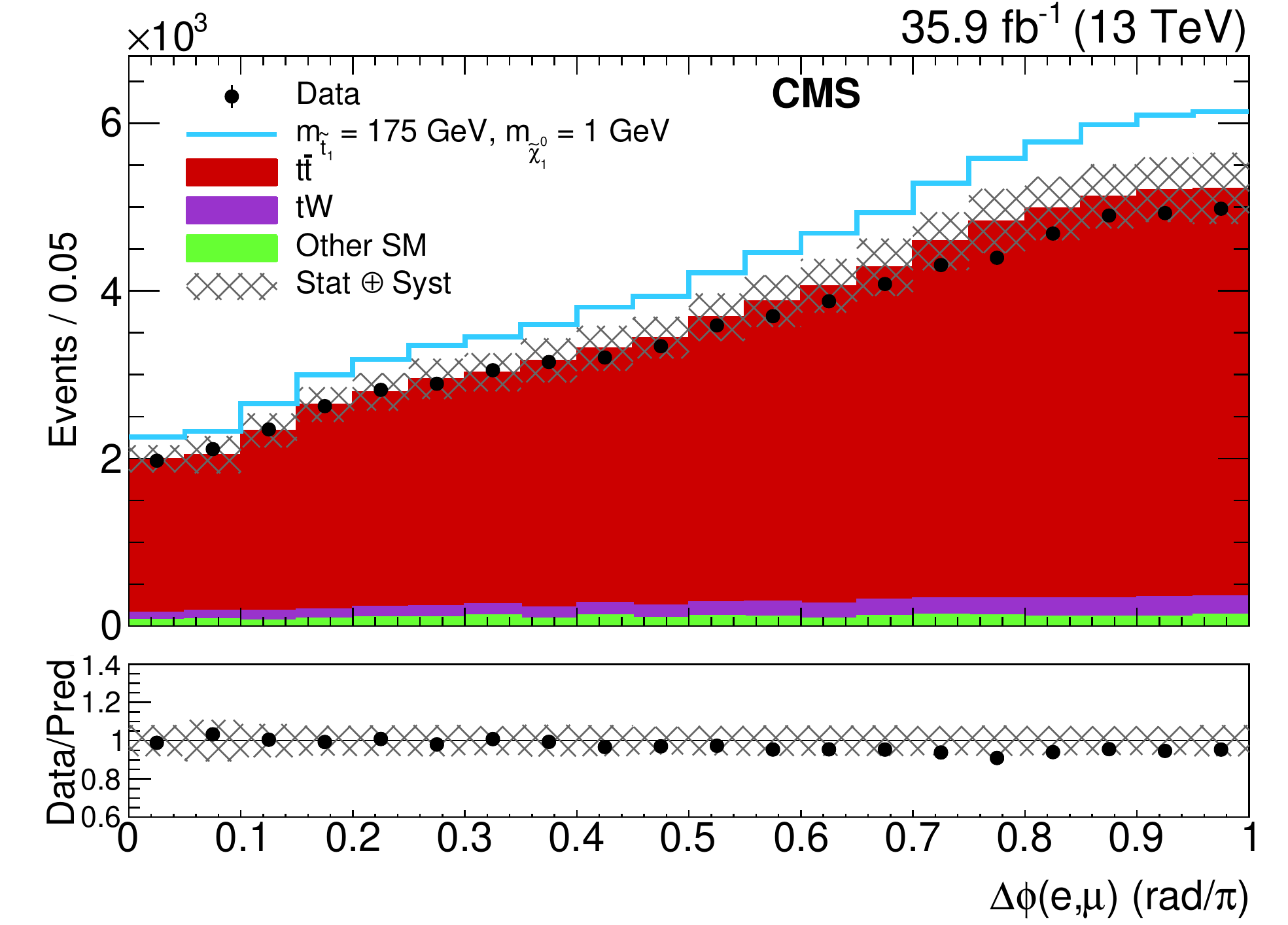}
     \includegraphics[width=1\cmsFigWidth]{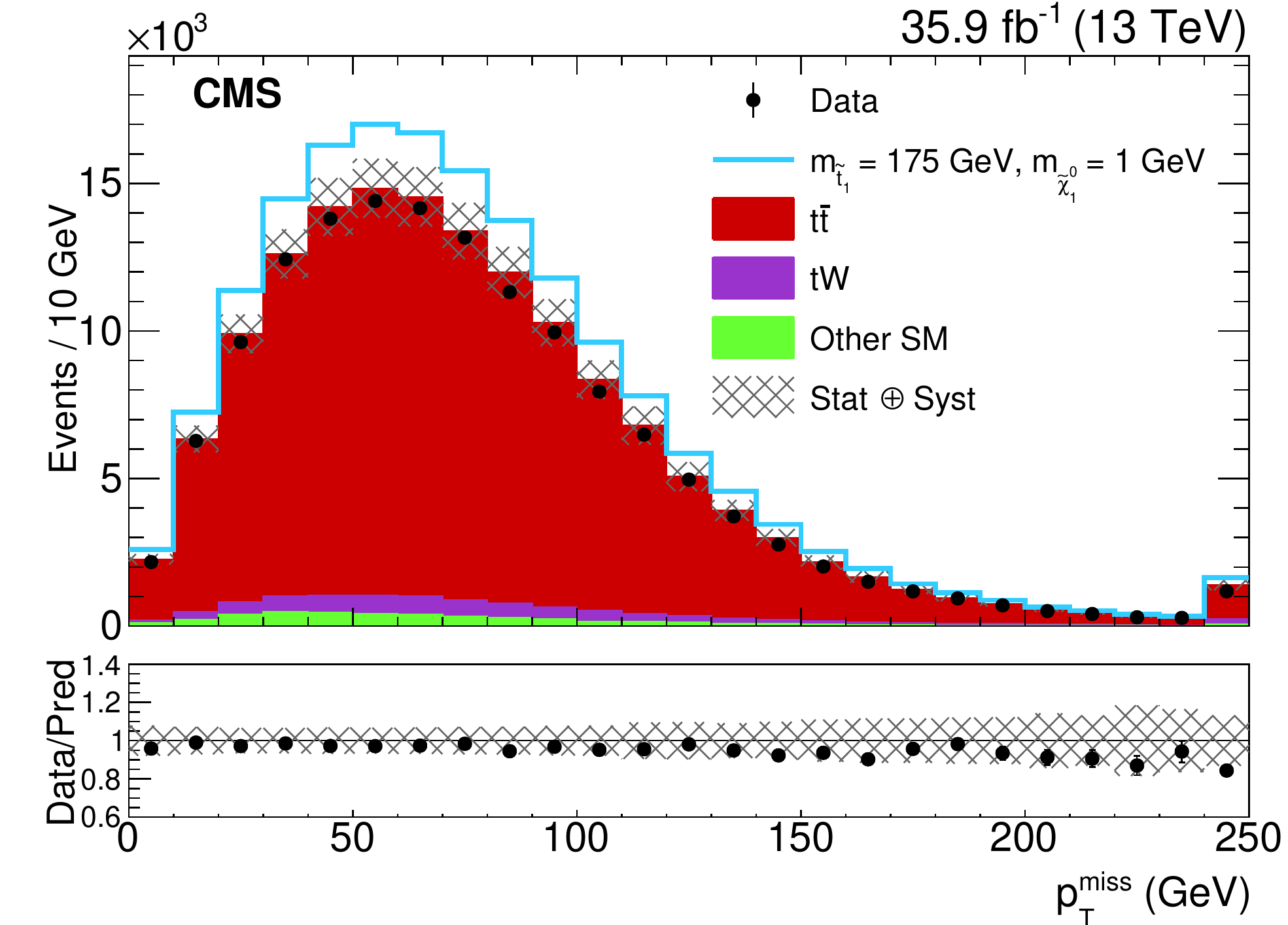}
     \caption{Distributions for leading and subleading lepton \pt, $\Delta\phi(\Pe,\Pgm)$, and \ptmiss. The uncertainty band includes statistical and all systematic uncertainties described in Section~\ref{sec:unc}. The last bin contains the overflow events. The signal is stacked on top of the background prediction for a mass hypothesis of $\mstop = 175\GeV$ and $\mchi = 1\GeV$.}
     \label{fig:controlplots}

 \end{figure}

A good agreement between data and SM predictions after the full event selection and after the corrections described in Section \ref{sec:selection} is observed, within the uncertainties, and is shown in Fig.~\ref{fig:controlplots} for the leading and subleading lepton \pt, \ptmiss, and the angle between the momentum of the leptons in the transverse plane ($\Delta\phi(\Pe,\Pgm)$). The considered uncertainties are described in Section~\ref{sec:unc}.

\section{Systematic uncertainties} \label{sec:unc}

Because of the large impact of the \ttbar~background prediction in this search, various modelling systematic uncertainties are assigned, reflecting the limited knowledge of the main theoretical parameters used in the simulation. The ranges of variation of these parameters were set in several previous CMS analyses \cite{CMS-PAS-TOP-16-021} and the modelling of the \ttbar background has been shown to accurately describe several kinematic variables within the systematic uncertainties~\cite{Sirunyan:2018ptc}.
 Details on the systematic uncertainties accounting for modelling effects are reported in Section \ref{sec:ttmodeling}.

The background and signal estimates are affected by several systematic uncertainties in the acceptance, efficiency, and normalization. The effect of uncertainties in the trigger efficiencies, lepton reconstruction, identification and isolation efficiencies, jet energy scale and resolution, pileup reweighting, and \cPqb~tagging efficiency and mistag rate efficiencies, are considered in the estimate of background and signal yields. These uncertainties are described in Section~\ref{sec:experimentalsyst}.

Some other uncertainties, including normalization uncertainties on \tW~and other minor backgrounds and modelling uncertainties on the signal, are described in Section~\ref{sec:otherunc}.

\subsection{Modelling uncertainties in the \ttbar background} \label{sec:ttmodeling}

An uncertainty of 6\% is assigned to the \ttbar background normalization, taking into account two effects. The first one is the uncertainty in the NNLO cross section from the variations in the PDFs, $\alpS$, and the scales calculated using the program \textsc{Top}++ for a top quark mass of 172.5\GeV~\cite{Czakon:2011xx}. The second effect is the uncertainty from the choice of the top quark mass obtained by varying it by $\pm$1\GeV in the calculation of the cross section.

In addition to the normalization uncertainty, several sources of modelling uncertainties are considered. All the modelling uncertainties are propagated to the \mtll shape and described in the next paragraphs. Their effect on the \ttbar yields is summarized in Table~\ref{tab:ttbarunc}.

The uncertainty in the modelling of the hard interaction process is assessed in the \POWHEG sample through changes of the $\mu_\mathrm{F}$ and $\mu_\mathrm{R}$ scales by factors of 2 and 1/2 relative to their common nominal value of $\mu_\mathrm{F}^2 = \mu_\mathrm{R}^{2} = \mtop^2 + p^2_{\mathrm{T},\cPqt}$. Here $p^2_{\mathrm{T},\cPqt}$ denotes the square of the transverse momentum of the top quark in the \ttbar rest frame.
The uncertainty in the choice of the PDFs and in the value of $\alpS$ is determined by reweighting the sample of simulated \ttbar events according to the envelope of a PDF set of 100 NNPDF3.0 replicas~\cite{Ball:2014uwa}. The uncertainty in $\alpS$ is propagated by reweighting the simulated sample by sets of weights with two variations within the uncertainties of $\alpS$.

The impact of the modelling uncertainties of the initial- and final-state radiation is evaluated by varying the parton shower scales (running \alpS) by factors of 2 and 1/2~\cite{bib:powheg2}. In addition, the impact of the matrix element (ME) and parton shower (PS) matching, which is parameterized by the \POWHEG generator as $h_{\mathrm{damp}} = 1.58^{+0.66}_{-0.59} \mtop$~\cite{CMS-PAS-TOP-16-021}, is calculated by varying this parameter within the uncertainties and propagating the result to the final yields.

The parameters of \PYTHIA~are tuned to model the measured underlying event~\cite{Skands:2014pea,CMS-PAS-TOP-16-021}. An uncertainty is assigned by varying these parameters within their uncertainties.

An uncertainty from the limited knowledge of the colour reconnection is estimated by comparing different models and taking as the uncertainty the maximum variation with respect to the nominal value for each bin. The procedure is described in detail in Ref.~\cite{Skands:2014pea}.

The top quark \pt in \ttbar events has been found to be slightly mismodelled~\cite{CMS-PAS-TOP-16-021}. A reweighting procedure, based on these studies, has been derived. To avoid biasing the search, the reweighting is not applied on the background estimate, but the difference between the weighted and unweighted distributions is taken as an uncertainty. The effect of the reweighting on the \ttbar yields is small and the range of the uncertainty can be seen in Table~\ref{tab:ttbarunc}.

A 1\GeV uncertainty in the top quark mass, which corresponds to twice the measured uncertainty by CMS \cite{Khachatryan:2015hba}, is also propagated to the acceptance. The differences in the \mtll yields for each bin of the distribution between the \ttbar background prediction with $\mtop = 172.5 \pm 1.0\GeV$ are taken as an uncertainty, accounting for the possible bias introduced in the choice of $\mtop = 172.5\GeV$ in the MC simulation.

\begin{table}[htb!]
 \centering \topcaption{Summary of the uncertainties on the \mtll distribution resulting from \ttbar background modelling uncertainties. The ranges correspond to variations of the uncertainty along the \mtll distribution. When only one number is shown, the uncertainty is approximately constant over the entire \mtll range.}
 \begin{tabular}{   l   c   }
 {                 Source } & {  Range    (\%)} \\
 \hline
 { $\mu_\mathrm{F}$ and $\mu_\mathrm{R}$ scales } &  0.3--1.0 \\
 {                    PDF } &  $\approx$0.6 \\
 {  Initial-state radiation} & 0.5--1.0 \\
 {  Final-state radiation} &  0.6--1.2\\
 { ME/PS matching ($h_{\mathrm{damp}}$) } &  0.3--2.0 \\
 {       Underlying event } &  $\approx$0.8 \\
 {     Colour reconnection } &  $\approx$1.5  \\
 {  Top quark \pt reweighting } &  0.1--0.5 \\
 {  Top quark mass (acceptance) } &  $\approx$1.0 \\
 \end{tabular}
 \label{tab:ttbarunc}
\end{table}

\subsection{Experimental uncertainties}\label{sec:experimentalsyst}
A summary of the effect of the experimental uncertainties on the \mtll distribution for events passing the full selection is shown in Table \ref{tab:expunc}.

The uncertainties in the dilepton trigger, lepton identification, and isolation efficiencies used in simulation are estimated by varying data-to-simulation scale factors by their uncertainties, which are about 1.5\% for electron and muon identification and isolation efficiencies, and about 0.5\% for the trigger efficiency.

To account for the uncertainties in the lepton momentum scales, the momenta of the leptons are varied by their uncertainties, which are of the order of 0.1--0.5\% for electrons~\cite{Khachatryan:2015hwa} and about 0.2\% for muons~\cite{Sirunyan:2018fpa}. The uncertainties associated with the jet energy scale and jet energy resolution are determined by varying these quantities in bins of \pt and $\eta$, according to the uncertainties in the jet energy corrections, which amount to a few percent.

The uncertainties associated with the \cPqb~tagging efficiency and mistag rate are determined by varying the scale factors for the \cPqb-tagged jets and mistagged light-flavour jets, according to their uncertainties, as measured in QCD multijet events~\cite{Sirunyan:2017ezt}. The average uncertainties on these scale factors for a \ttbar sample are of the order of 1.2\%, with a dependence on \pt and $\eta$.

The uncertainty in \ptmiss~from the contribution of unclustered energy is evaluated based on the momentum resolution of the different PF candidates, according to their classification. Details on the procedure can be found in Refs.~\cite{bib:PF,CMS:EGM-14-001,TRK-11-001}.

The uncertainty from the pileup reweighting procedure is evaluated by varying the inelastic \Pp\Pp~cross section by $\pm$4.6\%~\cite{Sirunyan:2018nqx}.

The uncertainty in the integrated luminosity, which affects the signal and background normalization, is estimated to be 2.5\%~\cite{CMS-PAS-LUMI-2017}.

\begin{table}[htb!]
 \centering \topcaption{Summary of the uncertainties in \ttbar~background and signal simulation resulting from experimental uncertainties. The numbers represent typical values of the uncertainties in the signal and \ttbar background yields or ranges for these uncertainties in different \mtll bins and in different signal samples.}
 \begin{tabular}{   l   c  }
 {         Source } & {  Range for \ttbar and signal (\%)} \\
 \hline
 {     Trigger efficiency }  & {$\approx$0.6} \\
 {  Muon efficiencies }  &  {$\approx$1.4} \\
 {  Electron efficiencies }  &  {$\approx$1.5} \\
 {    Lepton energy scale }  &  {0.5--2.0} \\
 {       Jet energy scale }  &  {1.5--3.0} \\
 {   Jet energy resolution}  &  {0.3--3.5} \\
 { \cPqb~tagging efficiency }  & {1.2--2.0} \\
 {      Mistag efficiency }  &  {0.2--0.6} \\
 {   Unclustered energy}  &  {0.5--1.5} \\
 {                Pileup }  &  {0.5--3.5} \\
 \end{tabular}
 \label{tab:expunc}
\end{table}

\subsection{Other uncertainties} \label{sec:otherunc}
A normalization uncertainty of 15\% is applied to the DY process, covering differences seen between data and MC predictions in different jet multiplicity regions~\cite{Khachatryan:2016kzg}. For other backgrounds, including \tW, dibosons, and {\ttbar}V, a normalization uncertainty of 30\% is assigned~\cite{Khachatryan:2016kzg}, covering the uncertainties in the predicted cross sections and possible extrapolation to the phase space used in the analysis. For the nonprompt lepton background, a normalization uncertainty of 30\% is applied, taking into account the effect of the limited number of MC events used in the estimation of the same-sign to opposite-sign transfer factor applied, and the normalization of the prompt-process subtraction in the control region.

Furthermore, a 15\% uncertainty in the signal normalization is assigned, according to the uncertainties in the predicted cross section of signal models in the top squark mass range of the analysis~\cite{Borschensky:2014cia}. The effect on the acceptance of the uncertainties in the factorization and renormalization scales is taken into account by varying $\mu_\mathrm{F}$ and $\mu_\mathrm{R}$ by factors of 2 and 1/2 both~\cite{Kalogeropoulos:2018cke}. This uncertainty is propagated to the signal yields, resulting in an uncertainty in each \mtll bins of the order of 0.5--1.0\%.

The \amcatnlo modelling of the initial-state radiation in signal events is improved by scaling the \pt distribution of the initial-state radiation jets in MC, according to a correction derived using \ttbar events, following the same procedure described in \cite{Sirunyan:2017leh}. An uncertainty is applied by considering variations of half the difference between the corrections and unity. The effect of this uncertainty on the signal yields amounts to about 1\%, with individual values assigned to each \mtll bin.

\section{Results}
\begin{figure}[htb!]
   \centering
     \includegraphics[width=1.5\cmsFigWidth]{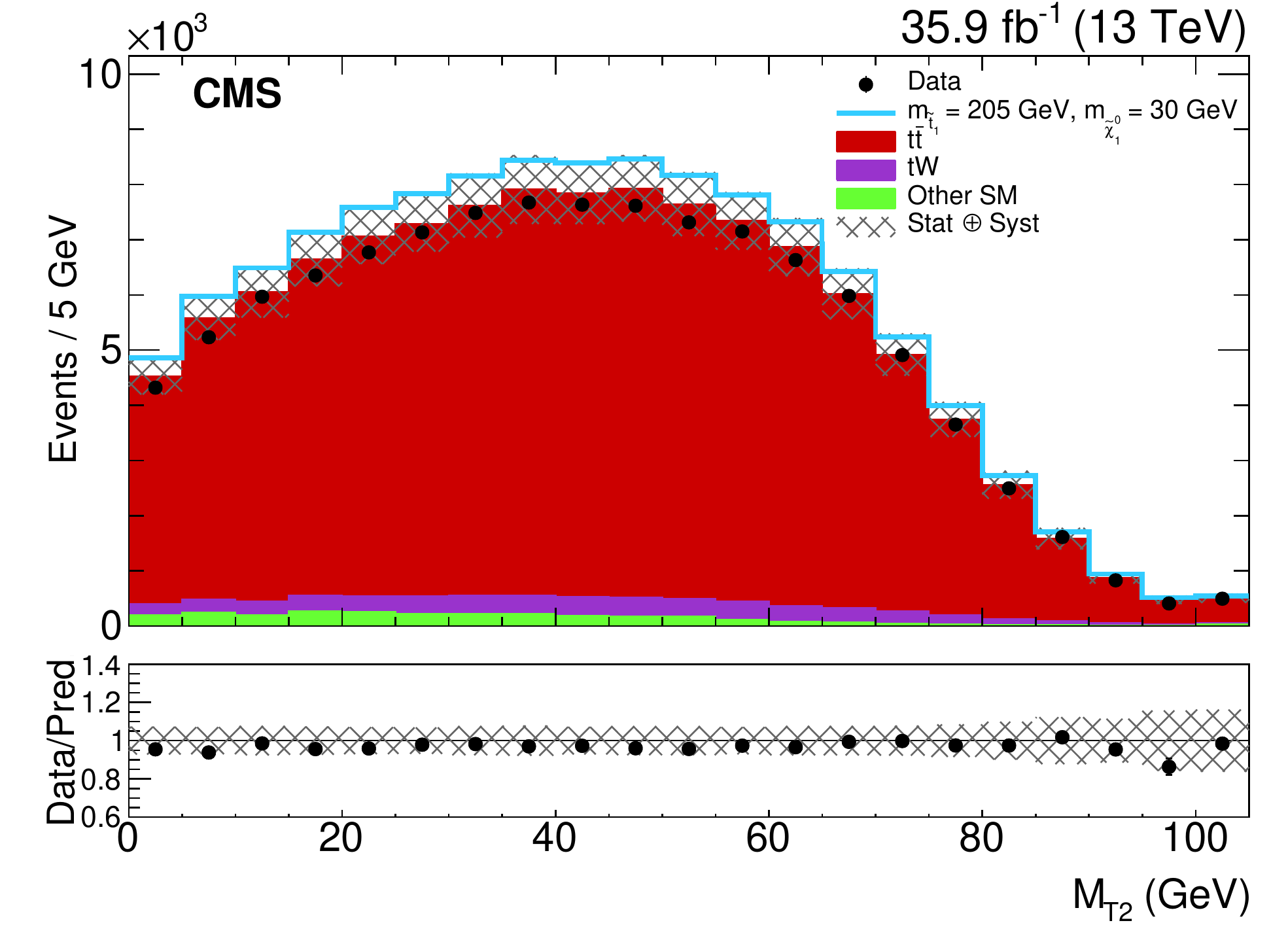}
     \caption{\mtll distribution (prefit) for data and predicted background. The \mtll distribution for a signal corresponding to a top squark mass of 205\GeV and a neutralino mass of 30\GeV is also shown, stacked on top of the background estimate. The hatched bands correspond to the combined systematic and statistical uncertainties on background rates. The last bin of the histogram includes the overflow events. The lower pane shows the ratio between the observed data and the predicted SM background.}
     \label{fig:mt2}

 \end{figure}

The predicted and observed \mtll distributions for selected events are shown in Fig.~\ref{fig:mt2}. No significant deviation from the SM expectation is observed. The integrated expected and observed number of events are shown in Table \ref{tab:yields}. The number of events with $\mtll > 90\GeV$ reflects the discriminating power for different top squark and neutralino masses at high values of \mtll.

\begin{table}[htb!]
 \centering \topcaption{Number of expected and observed events after the selection, with $\mtll > 0$ and \mbox{$\mtll > 90\GeV$}. The quoted uncertainties reflect both the statistical and systematic contributions.}
 \begin{tabular}{   l   c  c }
 {          Process } & { with $\mtll > 0\GeV$ } & {with $\mtll > 90\GeV$ }  \\
 \hline
 {               \ttbar } & $102\,400 \pm 7400$  & $1680 \pm 260$  \\
 {                   \tW } &  $4700 \pm 1400$  &  $92 \pm 32$ \\
 { Nonprompt leptons    } &  $1330 \pm 400 $  &  $30 \pm 11$ \\
 { DY + {\ttbar}V + Dibosons } &  $570  \pm 100 $  &  $19 \pm 6 $ \\[\cmsTabSkip]
 { Total Background } & $109\,000 \pm 7600$      & $1821 \pm 260$  \\[\cmsTabSkip]
 { Signal: $\mstop = 175.0\GeV$, $\mchi = 1.0\GeV$ } &  $16\,400 \pm 2500$        & $276 \pm 53$  \\
 { Signal: $\mstop = 205.0\GeV$, $\mchi = 22.5\GeV$ } &   $8070 \pm 1240$         & $232 \pm 41$  \\
 { Signal: $\mstop = 205.0\GeV$, $\mchi = 30.0\GeV$ } &   $7830 \pm 1200$         & $157 \pm 27$  \\
 { Signal: $\mstop = 205.0\GeV$, $\mchi = 37.5\GeV$ } &   $6140 \pm  650$         & $262 \pm 45$  \\
 { Signal: $\mstop = 242.5\GeV$, $\mchi = 67.5\GeV$ } &   $3550 \pm 540 $         & $106 \pm 19$  \\[\cmsTabSkip]
 {         Data } & 105\,893 & 1694 \\
 \end{tabular}
 \label{tab:yields}
\end{table}

The statistical interpretation is performed by testing the SM hypothesis against the SUSY hypothesis. A binned profile likelihood fit of the \mtll distribution is performed, where the nuisance parameters are modelled using log-normal distributions. All the systematic uncertainties described in Section~\ref{sec:experimentalsyst} and \ref{sec:ttmodeling} are assigned to each \mtll bin individually, and treated as correlated among all \mtll bins and all processes. The statistical uncertainties are treated as uncorrelated nuisance parameters in each bin of the \mtll distribution.

Upper limits on the top squark pair production cross section are calculated at 95\% confidence level (\CL) using a modified frequentist approach and the \CLs criterion, implemented through an asymptotic approximation~\cite{Cowan:2010js,Junk1999,Read:2002hq,CMS-NOTE-2011-005}. All the uncertainties in the background and signal predictions described in Section \ref{sec:unc} are modelled as nuisance parameters and profiled in the fit.

\begin{figure}[htb!]
   \centering
     \includegraphics[width=1\cmsFigWidth]{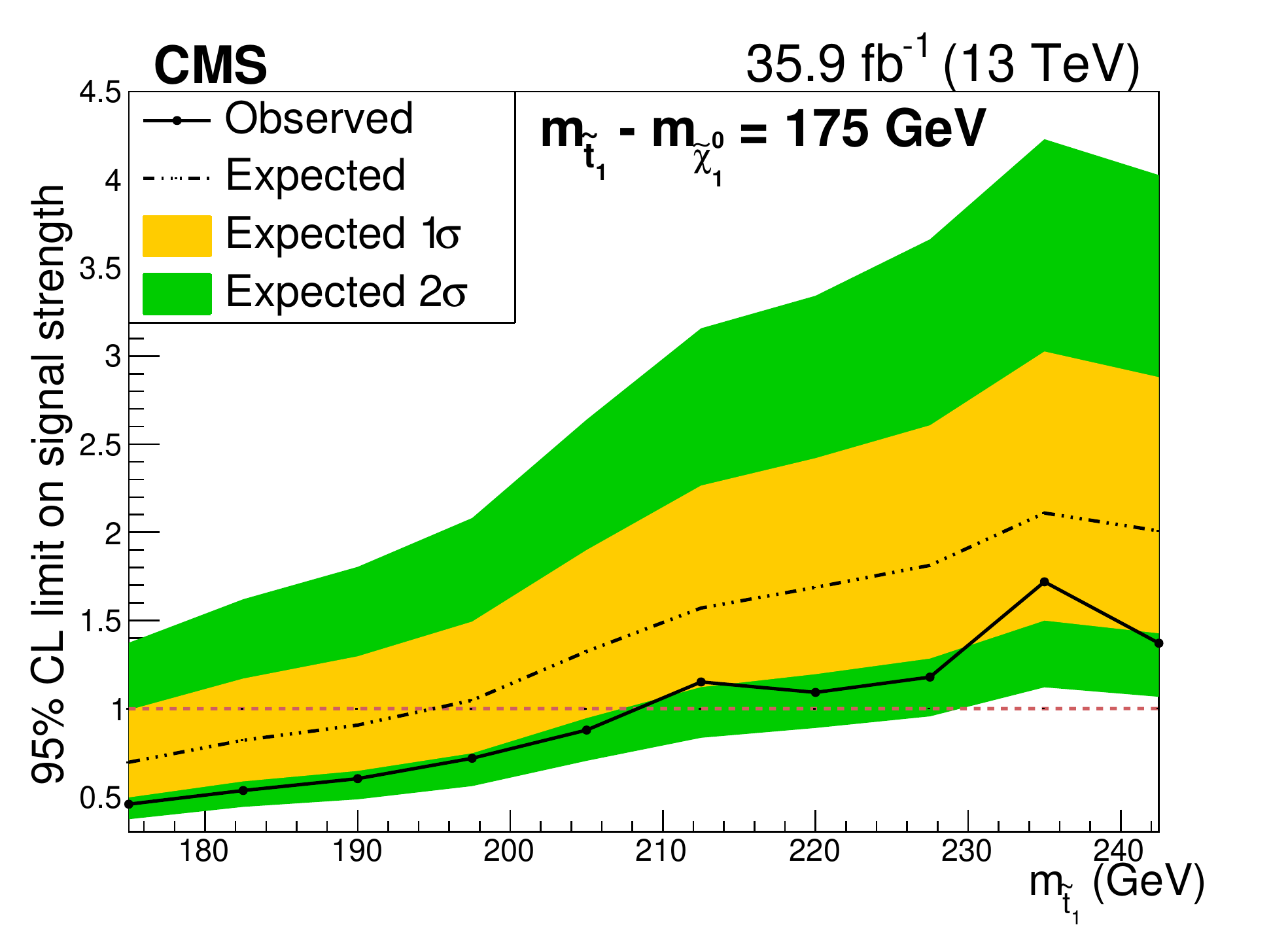}
     \includegraphics[width=1\cmsFigWidth]{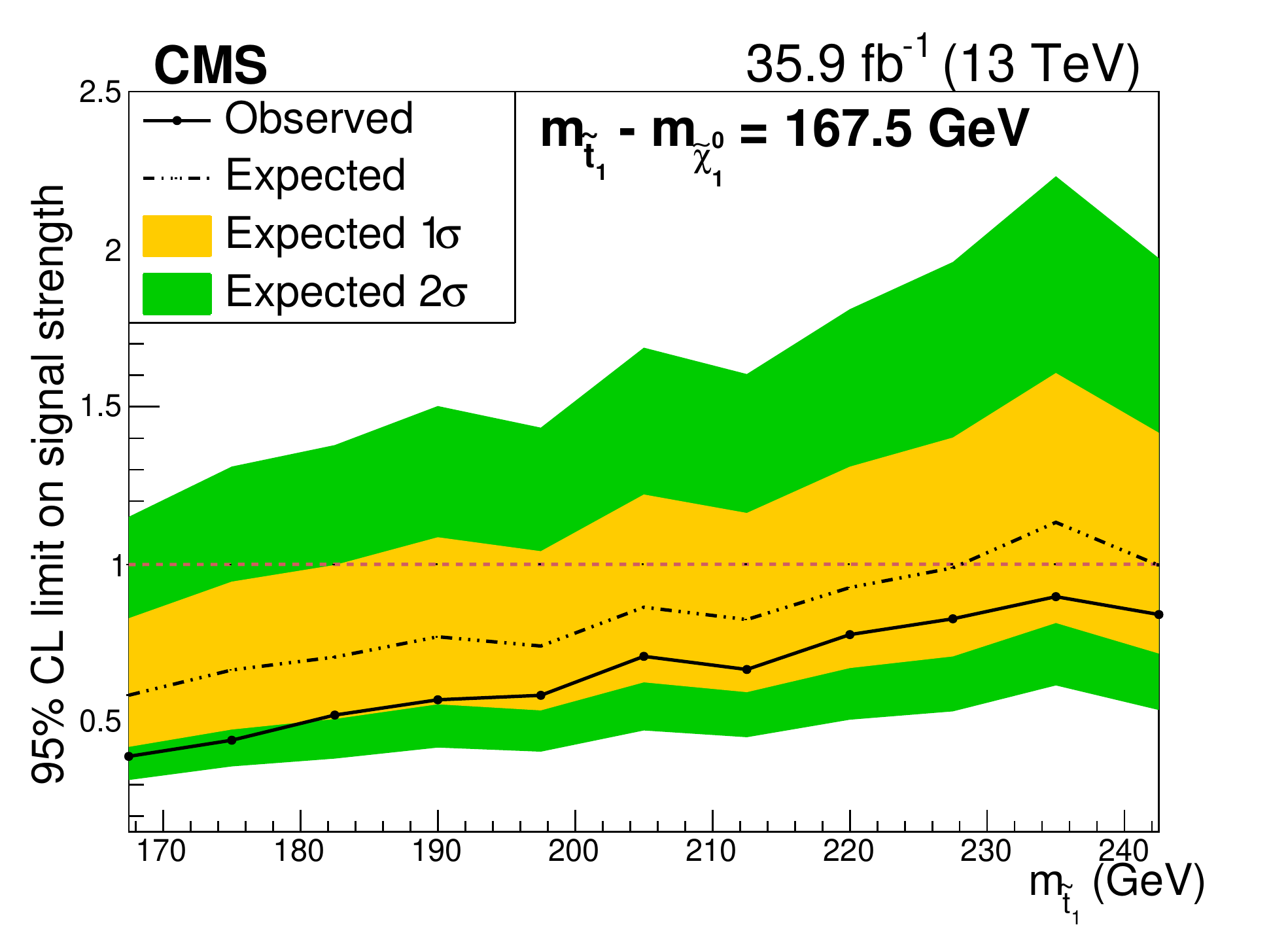}
     \includegraphics[width=1\cmsFigWidth]{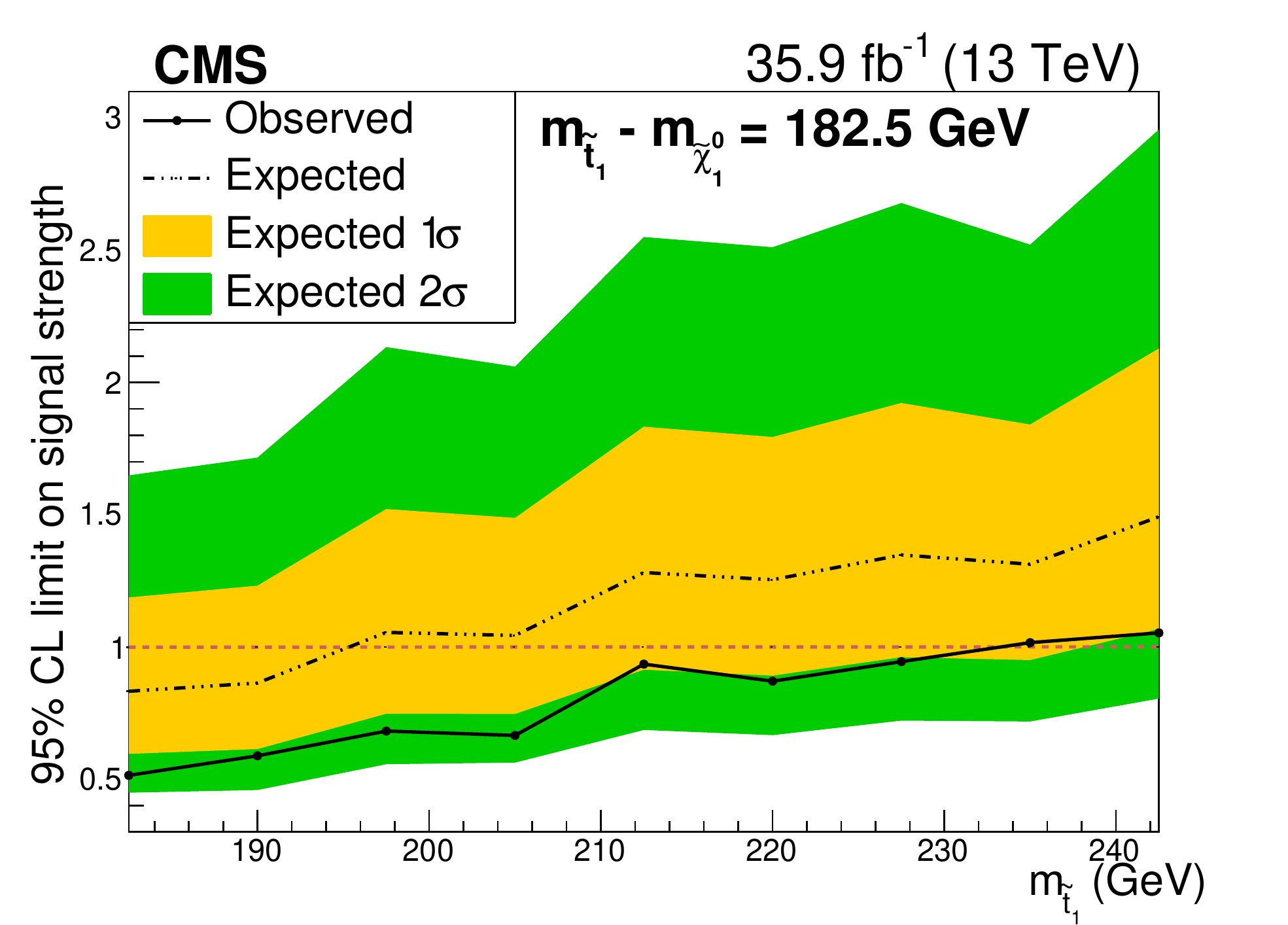}
     \caption{Expected and observed upper limits at 95\% \CL on the signal strength as a function of the top squark mass for $\mstop - \mchi = 175\GeV$ (upper left), $\mstop - \mchi = 167.5\GeV$ (upper right) and $\mstop - \mchi = 182.5\GeV$ (lower). The green dark and yellow light bands correspond to the 68 and 95\% \CL ranges of the expected upper limits.}
     \label{fig:brazil}

 \end{figure}

We interpret the results for different signals characterized by top squark masses from 170 to 250\GeV and by three different mass differences between the top squark and the neutralino: $\Delta m(\pstop, \pchi) = 167.5$, 175.0, and 182.5\GeV. The sensitivity of the analysis to SUSY models with low neutralino masses and $\Delta m(\pstop, \pchi) = \mtop$ comes mostly from the signal normalization, while the differences on \mtll shape become important for top squark masses greater than 210\GeV. For the difference in masses of $\Delta m(\pstop, \pchi) = 167.5$ and 182.5\GeV, the sensitivity of the analysis is mostly driven by the differences between the signal and \ttbar distributions for high \mtll values ($\mtll\gtrsim  80\GeV$). The expected and observed upper limits on the signal strength, defined as the ratio between the excluded and the predicted cross sections, are shown in Fig.~\ref{fig:brazil}.

We exclude the presence of a signal up to a top squark mass of 208\GeV for $\Delta m(\pstop,\pchi) - 175 = 0\GeV$ and up to top squark masses of 235 (242)\GeV for $\Delta m(\pstop,\pchi) - 175 =  {+}{(-)} 7.5$\GeV.

\section{Summary}
A search is presented for a top squark with a mass difference from the neutralino mass close to the top quark mass, $\mstop - \mchi \approx \mtop$, using events with one opposite-sign electron-muon pair, at least two jets, and at least one \cPqb~jet. The $\pstop\to\cPqt\pchi$ decay mode is considered, and different top squark masses are explored up to 240\GeV with neutralino masses of $\mchi \approx \mstop - \mtop$. The \mtll variable is used in a binned profile likelihood fit to increase the sensitivity, owing to the different kinematic distributions between the signal and the \ttbar~background. Further sensitivity is gained from the absence of a kinematic endpoint in this variable for the signal.

No excess is observed and upper limits are set at 95\% confidence level on the top squark production cross section for top squark masses up to 208\GeV in models with $\mstop - \mchi \approx \mtop$~and masses up to 235 (242)\GeV in models with a mass difference of ${+}{(-)} 7.5$\GeV. This result significantly extends the exclusion limits of top squark searches at the LHC to higher top squark masses in the region where $\mstop - \mchi \approx \mtop$, that was previously unexplored.

\begin{acknowledgments}
We congratulate our colleagues in the CERN accelerator departments for the excellent performance of the LHC and thank the technical and administrative staffs at CERN and at other CMS institutes for their contributions to the success of the CMS effort. In addition, we gratefully acknowledge the computing centres and personnel of the Worldwide LHC Computing Grid for delivering so effectively the computing infrastructure essential to our analyses. Finally, we acknowledge the enduring support for the construction and operation of the LHC and the CMS detector provided by the following funding agencies: BMBWF and FWF (Austria); FNRS and FWO (Belgium); CNPq, CAPES, FAPERJ, FAPERGS, and FAPESP (Brazil); MES (Bulgaria); CERN; CAS, MoST, and NSFC (China); COLCIENCIAS (Colombia); MSES and CSF (Croatia); RPF (Cyprus); SENESCYT (Ecuador); MoER, ERC IUT, and ERDF (Estonia); Academy of Finland, MEC, and HIP (Finland); CEA and CNRS/IN2P3 (France); BMBF, DFG, and HGF (Germany); GSRT (Greece); NKFIA (Hungary); DAE and DST (India); IPM (Iran); SFI (Ireland); INFN (Italy); MSIP and NRF (Republic of Korea); MES (Latvia); LAS (Lithuania); MOE and UM (Malaysia); BUAP, CINVESTAV, CONACYT, LNS, SEP, and UASLP-FAI (Mexico); MOS (Montenegro); MBIE (New Zealand); PAEC (Pakistan); MSHE and NSC (Poland); FCT (Portugal); JINR (Dubna); MON, RosAtom, RAS, RFBR, and NRC KI (Russia); MESTD (Serbia); SEIDI, CPAN, PCTI, and FEDER (Spain); MOSTR (Sri Lanka); Swiss Funding Agencies (Switzerland); MST (Taipei); ThEPCenter, IPST, STAR, and NSTDA (Thailand); TUBITAK and TAEK (Turkey); NASU and SFFR (Ukraine); STFC (United Kingdom); DOE and NSF (USA).

\hyphenation{Rachada-pisek} Individuals have received support from the Marie-Curie program and the European Research Council and Horizon 2020 Grant, contract No. 675440 (European Union); the Leventis Foundation; the A. P. Sloan Foundation; the Alexander von Humboldt Foundation; the Belgian Federal Science Policy Office; the Fonds pour la Formation \`a la Recherche dans l'Industrie et dans l'Agriculture (FRIA-Belgium); the Agentschap voor Innovatie door Wetenschap en Technologie (IWT-Belgium); the F.R.S.-FNRS and FWO (Belgium) under the ``Excellence of Science - EOS" - be.h project n. 30820817; the Ministry of Education, Youth and Sports (MEYS) of the Czech Republic; the Lend\"ulet (``Momentum") Program and the J\'anos Bolyai Research Scholarship of the Hungarian Academy of Sciences, the New National Excellence Program \'UNKP, the NKFIA research grants 123842, 123959, 124845, 124850 and 125105 (Hungary); the Council of Science and Industrial Research, India; the HOMING PLUS program of the Foundation for Polish Science, cofinanced from European Union, Regional Development Fund, the Mobility Plus program of the Ministry of Science and Higher Education, the National Science Center (Poland), contracts Harmonia 2014/14/M/ST2/00428, Opus 2014/13/B/ST2/02543, 2014/15/B/ST2/03998, and 2015/19/B/ST2/02861, Sonata-bis 2012/07/E/ST2/01406; the National Priorities Research Program by Qatar National Research Fund; the Programa Estatal de Fomento de la Investigaci{\'o}n Cient{\'i}fica y T{\'e}cnica de Excelencia Mar\'{\i}a de Maeztu, grant MDM-2015-0509 and the Programa Severo Ochoa del Principado de Asturias; the Thalis and Aristeia programs cofinanced by EU-ESF and the Greek NSRF; the Rachadapisek Sompot Fund for Postdoctoral Fellowship, Chulalongkorn University and the Chulalongkorn Academic into Its 2nd Century Project Advancement Project (Thailand); the Welch Foundation, contract C-1845; and the Weston Havens Foundation (USA).
\end{acknowledgments}

\bibliography{auto_generated}

\cleardoublepage \appendix\section{The CMS Collaboration \label{app:collab}}\begin{sloppypar}\hyphenpenalty=5000\widowpenalty=500\clubpenalty=5000\vskip\cmsinstskip
\textbf{Yerevan Physics Institute, Yerevan, Armenia}\\*[0pt]
A.M.~Sirunyan, A.~Tumasyan
\vskip\cmsinstskip
\textbf{Institut f\"{u}r Hochenergiephysik, Wien, Austria}\\*[0pt]
W.~Adam, F.~Ambrogi, E.~Asilar, T.~Bergauer, J.~Brandstetter, M.~Dragicevic, J.~Er\"{o}, A.~Escalante~Del~Valle, M.~Flechl, R.~Fr\"{u}hwirth\cmsAuthorMark{1}, V.M.~Ghete, J.~Hrubec, M.~Jeitler\cmsAuthorMark{1}, N.~Krammer, I.~Kr\"{a}tschmer, D.~Liko, T.~Madlener, I.~Mikulec, N.~Rad, H.~Rohringer, J.~Schieck\cmsAuthorMark{1}, R.~Sch\"{o}fbeck, M.~Spanring, D.~Spitzbart, W.~Waltenberger, J.~Wittmann, C.-E.~Wulz\cmsAuthorMark{1}, M.~Zarucki
% ...
\vskip\cmsinstskip
% ...
1:  Also at Vienna University of Technology, Vienna, Austria\\
% ...
\end{sloppypar}
\end{document}